\documentclass[aps,prl,reprint,superscriptaddress,showpacs]{revtex4-1}
\usepackage{amsmath,amssymb}
\usepackage{graphicx}
\usepackage[draft=false,dvipdfm,colorlinks,linkcolor=blue,citecolor=blue,urlcolor=blue,breaklinks]{hyperref}
\usepackage{hypcap}
\usepackage{lmodern}
\usepackage{textcomp}
\usepackage[T1]{fontenc}

\begin{document}
\title{Probing electron-electron interaction in quantum Hall systems\\with scanning tunneling spectroscopy}
\author{S. Becker}
\affiliation{II. Physikalisches Institut B and JARA-FIT, RWTH Aachen University, 52074 Aachen, Germany}
\author{C. Karrasch}
\affiliation{Institut f\"ur Theorie der Statistischen Physik and JARA-FIT, RWTH Aachen University, 52074 Aachen, Germany}
\author{T. Mashoff}
\author{M. Pratzer}
\author{M. Liebmann}
\affiliation{II. Physikalisches Institut B and JARA-FIT, RWTH Aachen University, 52074 Aachen, Germany}
\author{V. Meden}
\affiliation{Institut f\"ur Theorie der Statistischen Physik and JARA-FIT, RWTH Aachen University, 52074 Aachen, Germany}
\author{M. Morgenstern}
\affiliation{II. Physikalisches Institut B and JARA-FIT, RWTH Aachen University, 52074 Aachen, Germany}
\date{\today}
\begin{abstract}
Using low-temperature scanning tunneling spectroscopy applied to the Cs-induced two-dimensional electron system (2DES) on \textit{p}-type InSb(110), we probe electron-electron interaction effects in the quantum Hall regime. The 2DES is decoupled from bulk states and exhibits spreading resistance within the insulating quantum Hall phases. In quantitative agreement with calculations we find an exchange enhancement of the spin splitting. Moreover, we observe that both the spatially averaged as well as the local density of states feature a characteristic Coulomb gap at the Fermi level. These results show that electron-electron interaction can be probed down to a resolution below all relevant length scales.
\end{abstract}
\maketitle
Quantum Hall (QH) physics \cite{Klitzing} is a paradigm for the study of interacting quantum systems \cite{Prange}. In this respect, the III-V semiconductors are the very mature materials, although graphene catches up \cite{Andrei,*Bolotin2009,*Stroscio1,*Crommie}. The most intriguing QH phases are driven by electron-electron (e-e) interaction \cite{Stoermer, *Streifen, Skyrmion}, which, however, is screened by nearby gates and competes with disorder. Thus, a central challenge towards a microscopic investigation of QH physics dominated by e-e interaction is to provide a sufficiently clean and electrically decoupled system probed down to the relevant length scales, most notably the magnetic length $l_{\text{B}}=\sqrt{\hbar/(eB)} \simeq 10~\text{nm}$ ($6~\text{T}$). Scanning tunneling spectroscopy (STS) achieves the required nm resolution. Applying STS to an adsorbate induced 2D electron system (2DES) \cite{Aristov1994,Morgenstern2002}, some of us have shown that the states responsible for the integer QH transitions can indeed be probed with nm resolution \cite{Morgenstern2003-3,Hashimoto2008}. Theoretical analysis \cite{Florens} of these data and similar experimental results for graphite and graphene \cite{Stroscio2,*Nimii2009,*Luican2011} have also been published.

Here, we modify the 2DES in order to fully decouple it from the substrate and to reduce the disorder. This allows to probe e-e interaction effects. In particular, we observe an exchange enhancement (EE) of the spin splitting at odd filling factors in quantitative agreement with a parameter-free calculation. Moreover, we measure a Coulomb gap in the spatially averaged density of states (DOS) at the Fermi level $E_{\text{F}}$. This Coulomb suppression is in quantitative agreement with predictions for localized systems \cite{Pollak1970,Efros1975,Pikus1995}. Interestingly, we find a similar suppression in the \emph{local} DOS (LDOS), which is probably caused by fluctuation effects. Observing these hallmarks of the e-e interaction in STS is a crucial step towards a direct imaging of intriguing QH states such as stripe, bubble or fractional QH \cite{Stoermer,*Streifen,bubble} phases.

The home-built scanning tunneling microscope operates at $T=5~\text{K}$ in ultra-high vacuum (UHV) \cite{Mashoff2009}. The $dI/dV$ curves representing the LDOS of the sample are measured by lock-in technique at constant tip--surface distance stabilized at current $I_{\text{stab}}$ and voltage $V_{\text{stab}}$. A modulation voltage $V_{\text{mod}}$ is used to detect $dI/dV$ while ramping the sample voltage $V$.

\begin{figure}[tb]\capstart
\includegraphics[width=8.5cm]{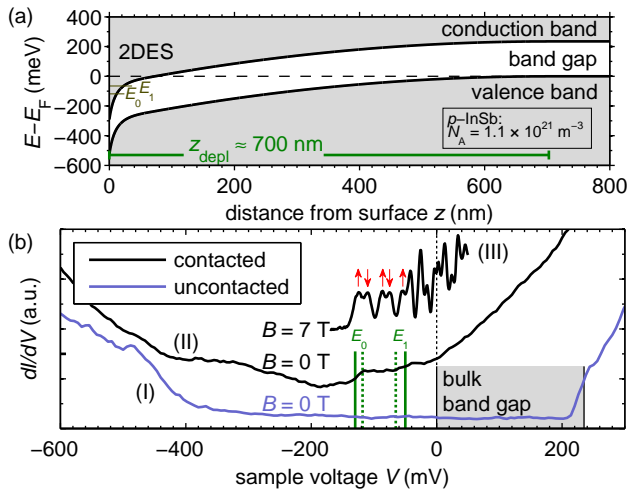}
\caption{(color online).
(a) Calculated band bending using 1D Poisson equation; first two 2D subband energies $E_0=-118~\text{meV}$, $E_1=-65~\text{meV}$ as calculated using the triangular well approximation \cite{Ando1984,supplement} are marked; corresponding electron distribution is shown in \cite{supplement}.
(b) Spatially averaged $dI/dV$ curves across an area $A_{\text{Av}}$: (I): without contacted 2DES, (II), (III): with contacted 2DES at $B = 0~\text{T}$, $7~\text{T}$ as marked; gray area: bulk band gap of \textit{p}-InSb \cite{Vurgaftman2001}; $E_{\text{0}}$, $E_{\text{1}}$: subband energies from experiment (solid) and calculation (broken lines); arrows in (III) mark spin split LLs.
(I) $V_{\text{stab}} = 400~\text{mV}$, $I_{\text{stab}} = 100~\text{pA}$, $V_{\text{mod}} = 1~\text{mV}_{\text{rms}}$,
$A_{\text{Av}}=200\ \times 160~\text{nm}^2$;
(II) $V_{\text{stab}} = 300~\text{mV}$, $I_{\text{stab}} = 50~\text{pA}$, $V_{\text{mod}} = 3~\text{mV}_{\text{rms}}$,
$A_{\text{Av}}=100 \times 100~\text{nm}^2$;
(III) $V_{\text{stab}} = 300~\text{mV}$, $I_{\text{stab}} = 200~\text{pA}$, $V_{\text{mod}} = 0.4~\text{mV}_{\text{rms}}$, $A_{\text{Av}}=300\ \times 300~\text{nm}^2$.
\label{fig:dIdV_spatial}}
\end{figure}

The 2DES was prepared in UHV by cleavage of a \textit{p}-type InSb single crystal ($N_{\text{A}} = 1.1 \times 10^{21}~\text{m}^{-3}$) and subsequent Cs adsorption of $1.1~\%$ of a monolayer ($3.7 \times 10^{16}~\text{m}^{-2}$) onto the cooled (110) surface \cite{supplement}.
The single Cs atoms act as surface donors, which bend the bands downwards and induce a 2DES \cite{Aristov1994,Betti2001,Hashimoto2008}.
Fig.~\ref{fig:dIdV_spatial}(a) shows the corresponding band bending in the near-surface region as calculated by a Poisson solver. This leads to a 2DES with density $N_{\text{s}}\simeq 2.7 \times 10^{16}~\text{m}^{-2}$.
Note that the band bending reaches deep into the bulk leading to a decoupling of the confined states of the 2DES at $E_{i}$, $i\in\mathbb{N}_{0}$, from the partly empty bulk valence band (BVB) $600~\text{nm}$ apart.
Indeed, the spatially averaged $dI/dV$ curve (I) of Fig.~\ref{fig:dIdV_spatial}(b), measured without contacting the 2DES directly, does not exhibit any signature of the 2DES, but only an increase in $dI/dV$ close to the onset of the bulk conduction band ($+200~\text{mV}$) and the surface valence band ($-400~\text{mV}$).
The tunneling path from the 2DES to the BVB is blocked.
This is in contrast to measurements using \textit{n}-type \cite{Morgenstern2002,Hashimoto2008,Kanisawa2001} and \textit{p}-type samples with higher doping \cite{Wiebe2003,Becker2010}, always exhibiting a step like increase in spatially averaged $dI/dV$ curves close to the calculated $E_{i}$.
If our 2DES is additionally contacted by an Ag stripe running perpendicular to the cleavage plane \cite{Masutomi2007}, it exhibits two steps close to the calculated $E_{0}$ and $E_{1}$ as visible in curve (II) of Fig.~\ref{fig:dIdV_spatial}(b).

Applying a magnetic field $B$ perpendicular to the 2DES results in peaks corresponding to Landau levels (LLs) and spin levels of the 2DES [curve (III), Fig.~\ref{fig:dIdV_spatial}(b)]. Their distance is in accordance with the effective mass $m^{*}$ and $g$-factor $g^{*}$ of the InSb conduction band \cite{supplement}. The width of the peaks at lower energy is $\Delta E \simeq 16~\text{meV}$ which is caused by the potential disorder, mainly given by the dopants of the substrate. The Cs atoms, which are ionized by only $30~\%$ \cite{Getzlaff2001}, have a minor effect \cite{Morgenstern2002}.

\begin{figure}[tb]\capstart
\includegraphics[width=8.5cm]{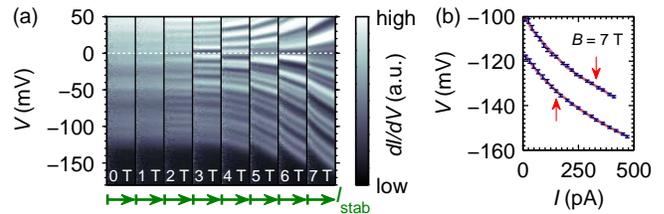}
\caption{(color online).
(a) $dI/dV$ spectra (grayscale) recorded at the same lateral position at different $B$ fields as indicated. $I_{\text{stab}}$ is increased for each $B$ from $100~\text{pA}$ to $2000~\text{pA}$ (left to right). $V_{\text{stab}}=300~\text{mV}$, $V_{\text{mod}} = 1.6~\text{mV}_{\text{rms}}$.
(b) Measured lowest LL positions (spin up $\uparrow$; spin down $\downarrow$) at $B=7~\text{T}$ (symbols) in comparison with calculated LL positions (line). Note that the current $I$ at the peak position and not $I_{\text{stab}}$ is used as $x$ axis.
\label{fig:Imod}}
\end{figure}

Further evidence for the electrical decoupling of the 2DES from the BVB is presented in Fig.~\ref{fig:Imod}, where $dI/dV$ curves at increasing $I_{\text{stab}}$ are shown for different $B$. All curves are measured at the same lateral position. With increasing $B$, pairs of lines corresponding to spin split LLs appear. For $B=0~\text{T}$, increasing $I_{\text{stab}}$ does not change the $dI/dV$ spectra. At higher $B$ the spectra are spread in $V$ with increasing $I_{\text{stab}}$. At $7~\text{T}$, the spin splitting of the lowest LL is increased by 13~\% and the LL distance is increased by 18~\%. The spreading is symmetric around $V=0~\text{mV}$, i.e.\ around $E_{\text{F}}$. It is attributed to the increased localization of electrons with growing $B$ leading to a decrease of 2D conductivity \cite{Prange}. The spreading cannot be explained by tip induced band bending (Stark effect) \cite{Feenstra1987,*Dombrowski1999}, which might increase with $B$ due to a reduced screening of the 2DES. Poisson calculations reveal that the spreading at largest tip--surface distance must be much larger than the spreading induced by the change in tip--surface distance in contrast to experiment. Instead, the spreading is quantitatively reproduced by assuming a thermally activated nearest neighbor hopping of the localized electrons within the 2DES from or towards the tip. The model uses barrier heights and next-neighbor distances of valleys as determined from spatially resolved $dI/dV$ data \cite{supplement} and assumes a reasonable attempt frequency of $\nu_0=10^{13}$~Hz \footnote{Details will be subject of a forthcoming publication.}. The resulting peak positions in comparison with the experimental data for $B=7~\text{T}$ are shown in Fig.~\ref{fig:Imod}(b). The excellent agreement strongly supports our assumption that the current indeed flows along the 2DES exhibiting reduced conductivity with increasing $B$.

\begin{figure}[tb]\capstart
\includegraphics[width=8.5cm]{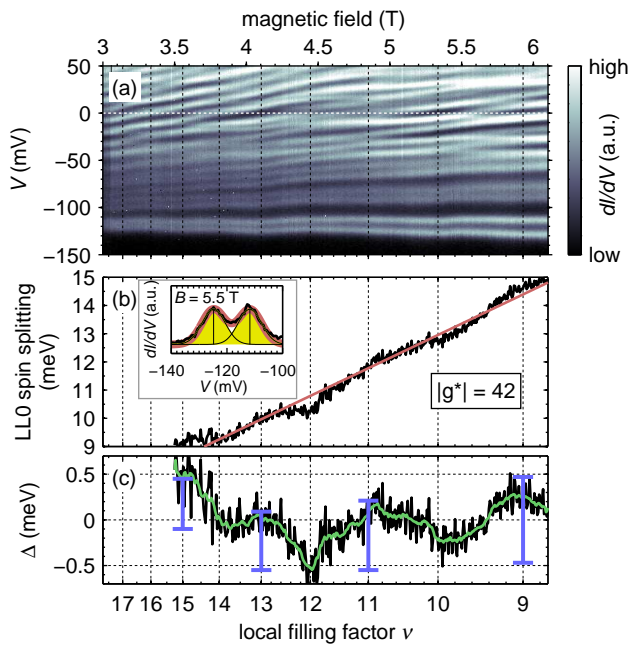}
\caption{(color online).
(a) Landau fan showing $dI/dV$ as grayscale; $B$ field ramped downwards, $V_{\text{stab}}=300~\text{mV}$, $I_{\text{stab}}=400~\text{pA}$, $V_{\text{mod}} = 1.6~\text{mV}_{\text{rms}}$.
(b) Spin splitting energy $E_{\text{SS}}$ of lowest LL extracted by Gaussian fits as shown in the inset. Straight line marks $E_{\text{SS}}=|g^{*}|\mu_{\text{B}} B$.
(c) Deviation $\Delta$ from the linear fit in (b). The inner bright line is smoothed. Vertical bars show the calculated values of EE between neighboring odd and even filling factors. Local filling factors $\nu \propto 1/B$ are matched by counting spin levels below $E_{\text{F}}$ (compare Fig.~\ref{fig:CoulombGap}(b)) and verifying their $B$ dependence.\label{fig:SpinSplitting}}
\end{figure}

The surface 2DES, thus, is occupied, exhibits Landau as well as spin quantization, has moderate disorder, and is decoupled from the bulk electrons of InSb. Moreover, the center of mass of the 2DES is 8~nm below the surface and, thus, sufficiently far from the metallic tip to prevent complete screening. These are the requirements to observe e-e interaction effects within the QH regime. One such effect is the EE of the spin splitting. Loosely speaking the effective repulsion between electrons with parallel spins is smaller than the one for antiparallel spins. This eventually leads to an increase of the spin splitting energy $E_{\text{SS}}$ at odd filling factors \cite{Ando1974}. Fig.~\ref{fig:SpinSplitting}(a) shows $dI/dV$ spectra taken at a fixed position while ramping the magnetic field $B$ providing the so-called Landau fan.
Less than 10~\% of the fanning is caused by the spreading resistance described above. Varying $B$, the conductance lines are wavy and deviate from
$E^{n}_{i,\pm}=E_{i} + \hbar\omega_{\text{c}} (n+\frac{1}{2}) \pm \frac{1}{2} g^{*} \mu_{\text{B}} B$
with subband index $i$, LL index $n\in\mathbb{N}_{0}$, spin index $\pm$, cyclotron frequency $\omega_{\text{c}} = {eB}/{m^{*}}$, and Bohr magneton $\mu_{\text{B}}$. One obvious reason for waviness is a shift of $E_{\text{F}}$ with magnetic field taking place once the increasing degeneracy of a LL favors a transition to the next LL. More importantly, $g^{*}$ is filling factor dependent due to the EE. To analyze this in more detail, we concentrate on the lowest LL around $-120~\text{mV}$, which gives the highest accuracy in determining $E_{\text{SS}}$.
We adapted two Gaussians for all 386 spectra between $3.5$ and $6.1~\text{T}$ \footnote{Gaussians of equal width and height were fitted using a nonlinear least squares method and a trust-region algorithm as implemented in Matlab, see \href{http://www.mathworks.com/help/toolbox/curvefit/}{MathWorks Curve Fitting Toolbox V2.1 User's Guide}}. The fits are good as can be seen in the inset of Fig.~\ref{fig:SpinSplitting}(b) and by the confidence value of $R^2=0.94$ ($0.97$ above $5~\text{T}$). The error for the resulting $E_{\text{SS}}$ is about $0.2~\text{meV}$. $E_{\text{SS}}(B)$ is shown in Fig.~\ref{fig:SpinSplitting}(b) in comparison to a straight line corresponding to ordinary Zeeman splitting of $|g^{*}| \mu_{\text{B}} B$ with $|g^{*}|=42$. Fig.~\ref{fig:SpinSplitting}(c) shows the deviation $\Delta(B)$ from the straight line. It oscillates around 0~meV with maxima (minima) around odd (even) filling factors as expected for EE \cite{Janak1969}.  The not expected negative values of $\Delta(B)$ are probably caused by slight deviations from a spin splitting linear in $B$ due to either increasing spreading with $B$, which leads to superlinearity, or nonparabolicity of InSb leading to a smooth decrease of $g^{*}$, thus, supralinearity. However, both effects cannot explain oscillations in $\Delta(B)$. One could imagine that spreading depends also on filling factor being largest at even ones, but that would lead to an oscillation with maxima at even filling factor.

Moreover, the amplitude of the $\Delta(B)$ oscillation is about $0.7~\text{meV}$ in excellent agreement with theoretical estimates for EE [vertical bars in Fig.~\ref{fig:SpinSplitting}(c)]. They are obtained by treating the Coulomb interaction using a random phase approximation. This is well justified since the subband electron density $N_{0}$ is large compared to the scale set by the Bohr radius \cite{Ando1974,MacDonald1992,supplement}. We performed the calculation using $m^{*}=0.02m_{0}$ and $g^{*}=-42$ but emphasize that the results barely change if these or other system parameters are varied within reasonable limits (e.g.\ less than 1~\% for $g^{*}=-38$). Thus, magnitude and oscillation phase of $\Delta(B)$ compare favorably with a parameter-free calculation of EE. This implies that the \emph{short-ranged} e-e interaction effect EE can be probed by STS.

\begin{figure}[bt]\capstart
\includegraphics[width=8.5cm]{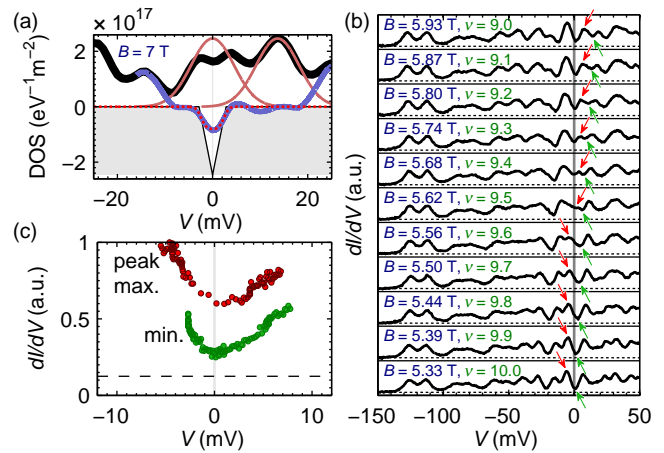}
\caption{(color online).
(a) Zoom of Fig.~\ref{fig:dIdV_spatial}(b)(III) (thick line), additional Gaussian fits for the LL at $E_{\text{F}}$ (thin lines), and a plot of the difference (medium, blue line) in comparison with the expected bare Coulomb gap of eq. {\protect \ref{eq:CoulombGap}} (V-shaped line) and the Coulomb gap taking finite temperature, screening and energy resolution into account (dotted line); absolute scale of DOS is deduced by fitting the lowest spin-split LL to a two-peak Gaussian and matching the integral of one Gaussian to the known level degeneracy $eB/(2\pi\hbar)$.
(b) $dI/dV$ spectra taken at the same position at $B$ as marked. Arrows of same color follow the same peak or minimum across $E_{\text{F}}$; a movie of the data is available in {\protect \cite{supplement}} (c) $dI/dV$ value of the marked peak and minimum in (b) as a function of energy (voltage).
\label{fig:CoulombGap}}
\end{figure}
For localized electrons interacting via the \emph{long-ranged} part of the Coulomb repulsion, the averaged tunneling DOS is expected to show a gap at $E_{\text{F}}$ \cite{Pollak1970,Efros1975,Pikus1995}. For a 2DES with unscreened repulsion at $T=0$~K, a qualitative analysis gives \cite{Efros1975,Pikus1995}
\begin{equation}
D_{0}(E)=\frac{2}{\pi}\frac{(4\pi\varepsilon_{0}\varepsilon_{\text{r}})^{2}}{e^4}\left|E-E_{\text{F}}\right|~.\label{eq:CoulombGap}
\end{equation}
More elaborate analytical and numerical results leave no doubt about the existence of a Coulomb gap while the exact shape remains controversial \cite{Pikus1995,Glatz}. This is due to the underlying (spin-)glass physics \cite{Glatz} known to be notoriously complex. A linear Coulomb gap was deduced from various experiments \cite{Butko,*Ashoori,*Ashoori2,*Hansen}. In our case, where the ratio between disorder and e-e interaction is
$R= \Delta E /[(e^2 \sqrt{N_{\text{s}}})/(4\pi \varepsilon_{\text{r}} \varepsilon_{0})]\simeq 1.1$,
we also find a dip in the DOS at $E_{\text{F}}$. Fig.~\ref{fig:CoulombGap}(a) shows the \emph{spatially averaged} $dI/dV$ curve (thick line) at $B=7~\text{T}$. Instead of a peak at $E_{\text{F}}$, one observes a double-peak with a minimum at 0~mV. The sum of two identical Gaussian peaks (thin lines)---mimicking the two spin levels of this particular LL (see \cite{supplement})---matches the measured DOS except of a suppression at $E_{\text{F}}$. Taking the difference between measured DOS and the sum of the two Gaussians eliminates all single-particle effects leaving only the dip at $E_{\text{F}}$ (medium line). If we modify Eq.~(\ref{eq:CoulombGap}) to account for finite temperature and screening effects \cite{Pikus1995} as well as for the energy resolution of our experiment of $1.6~\text{meV}$ \cite{supplement}, we obtain the dotted curve in Fig.~\ref{fig:CoulombGap}(a). It shows excellent agreement with the measured dip. The screening is taken to be caused by the STM tip being 8.6 nm away from the center of mass of the 2DES \cite{supplement}. Note that we observe the gap even around the critical state, i.e., close to half filling of a spin-polarized LL, which is consistent with numerical studies \cite{MacDonald1993}. The facts that we do not observe the dip at $E_{\text{F}}$ without localization (at $B=0~\text{T}$) and that we can reproduce it by a reasonable, parameter-free calculation strongly suggests that we observe the Coulomb gap. We can rule out inelastic excitations as a cause which would lead to much larger half widths of the gap (optical phonons: 22 meV, plasmons: 60~meV, spin excitations: 18~meV) and we are not aware of any many-particle mechanism besides the long-ranged Coulomb repulsion of localized electrons leading to a gap with the observed characteristics.

Surprisingly, a Coulomb gap---although typically thought of being a phenomenon related to disorder averaging or spatial averaging---is also observed in the \emph{local} DOS \cite{Morgenstern2002-2}. The intensity of a particular LDOS peak is suppressed when moved through $E_{\text{F}}$ by increasing $B$. Fig.~\ref{fig:CoulombGap}(b) shows corresponding $dI/dV$ curves at fixed position. The upper arrows follow a single spin level as it crosses $E_{\text{F}}$ and the peak intensity is plotted in Fig.~\ref{fig:CoulombGap}(c). A minimum intensity is observed exactly at $E_{\text{F}}$ ($B=5.62~\text{T}$) where the peak is suppressed by $48~\%$ (suppression in averaged DOS: $33~\%$). The same kind of $dI/dV$ suppression is found for the minimum between LLs, which is marked by the lower lying arrows in (b) and plotted in (c), too. A $dI/dV$ suppression at $E_{\text{F}}$ is also found for fixed $B$, if different positions are probed within the potential landscape \cite{supplement}. The finding of Coulomb suppression in the LDOS requires further studies and might be related to Coulomb glass dynamics \cite{Efros2001}.

In summary, we have shown that low-temperature STS is able to detect e-e interaction in QH samples down to a resolution below all relevant length scales. We have found an exchange enhancement (EE) of the spin splitting at odd fillings and a Coulomb suppression of the averaged as well as of the local DOS at $E_{\text{F}}$. The EE is in quantitative agreement with a well justified theory, while, due to the less clear status of theory, the comparison for the Coulomb gap is with calculations based on qualitative arguments only. No well-developed theory for the LDOS exists and we conjecture that the Coulomb gap in LDOS is related to (spin-)glass physics.

We acknowledge support by the DFG (MO 858/11-1).

\renewcommand{\thefigure}{S\arabic{figure}}
\renewcommand{\theequation}{S\arabic{equation}}
\setcounter{figure}{0}
\setcounter{equation}{0}
\section{Supplementary Information}
\section{Experimental details}
\begin{figure}\capstart
\includegraphics[width=8.5cm]{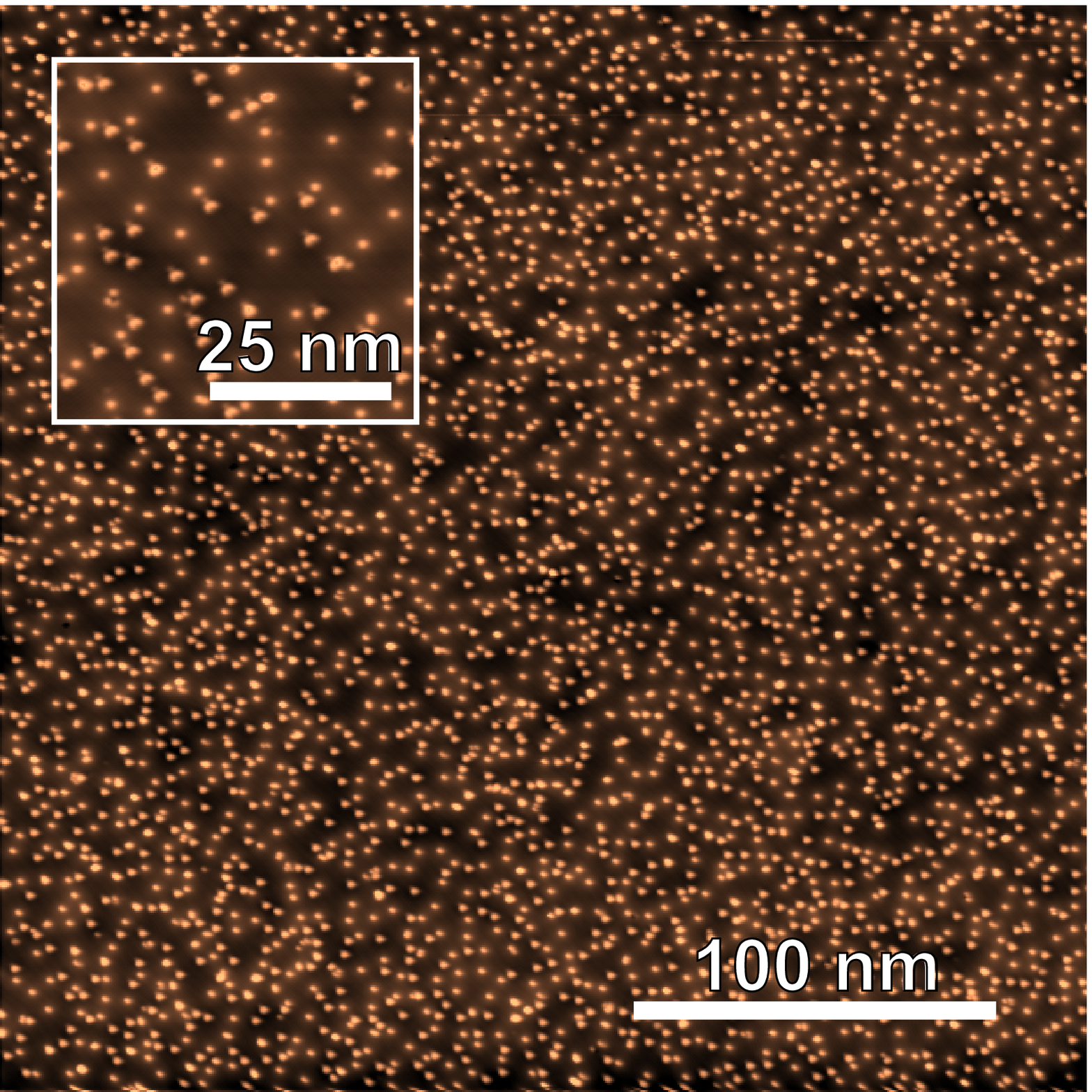}
\caption{STM image of InSb(110) with 1.1~\% Cs coverage; $V = 300~\text{mV}$, $I = 30~\text{pA}$, $300~\text{nm} \times 300~\text{nm}$. The inset shows an area of $50~\text{nm} \times 50~\text{nm}$ from a separate measurement with $V = 300~\text{mV}$, $I = 50~\text{pA}$.
\label{fig:topography}}
\end{figure}

For sample preparation, we glued a \textit{p}-InSb single crystal of size $3~\text{mm} \times 3~\text{mm} \times 3~\text{cm}$ to a sample holder using silver epoxy. A silver epoxy line drawn at one side of the crystal from the sample holder towards the surface serves as a direct electrical contact to the 2DES. At the opposite side of the epoxy line, the crystal was cut for controlled cleavage about $1~\text{mm}$ in depth. This cut is parallel to the sample holder surface and located about $1\text{--}2~\text{mm}$ above the sample holder. After bake-out within the load-lock of our ultrahigh vacuum system, the crystal was cleaved at room temperature and a pressure of $p\simeq 10^{-8}~\text{Pa}$ exhibiting the (110) surface. Then, we transferred the crystal into the pre-cooled scanning tunneling microscope (STM). For this purpose, the STM was lifted from the bath cryostat to the transfer chamber without breaking the vacuum \cite{Mashoff2009}. Cesium from a well-outgassed dispenser (SAES Getters) was evaporated onto the cold crystal surface. During and after the adsorption process the sample was held below $50~\text{K}$ to prevent diffusion induced Cs clustering \cite{Becker2010}. Fig.~\ref{fig:topography} shows a resulting STM image. The image area is atomically flat and each bright dot corresponds to a single cesium atom or, partly, to a Cs dimer which both act as donors. 3300 adsorbates are visible which corresponds to a coverage of $1.1~\%$ per InSb(110) surface unit cell.

The STM tip was etched from a tungsten wire outside of the vacuum system and prepared within the STM by field emission and consecutive voltage pulses on a W(110) crystal as well as on the InSb(110) surface itself. After such voltage pulses on InSb, the scan area had to be changed using the translation stage of the STM. We used tungsten because it enabled us to give good results on a similar sample system not exhibiting tip-induced band bending \cite{Becker2010}.

\section{Landau levels and potential disorder}
\begin{figure}\capstart
\includegraphics[width=8.5cm]{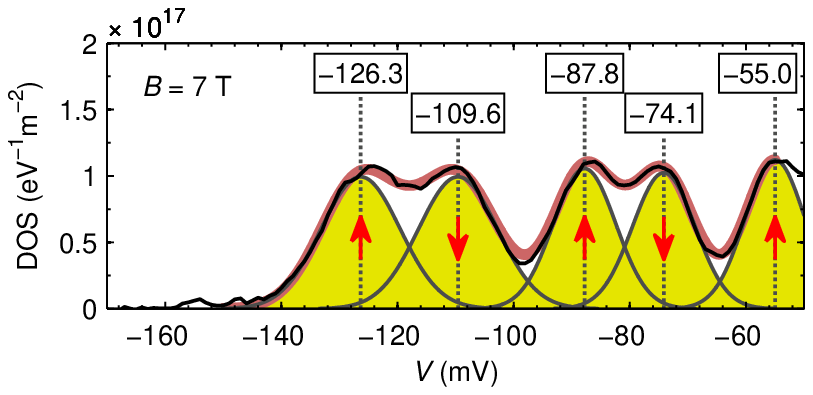}
\caption{Spatially averaged $dI/dV$ curve (black line) at $B = 7~\text{T}$, $V_{\text{stab}} = 300~\text{mV}$, $I_{\text{stab}} = 200~\text{pA}$, $V_{\text{mod}} = 0.4~\text{mV}_{\text{rms}}$, $A_{\text{Av}}=300~\text{nm} \times 300~\text{nm}$, $20\times 20$ spectra; yellow areas show the individual Gaussian fit curves of the spin split Landau levels marked by arrows each; red line marks the resulting fit of the $dI/dV$ curve gained by adding up the Gaussians; peak energies of the Gaussians are labeled above each peak in meV.\label{fig:dIdV_spatial2}}
\end{figure}

At $B=7~\text{T}$, we observe the spin split Landau levels of the first subband as shown in Fig.~\ref{fig:dIdV_spatial}(b) of the main article. The same measurement is shown in Fig.~\ref{fig:dIdV_spatial2} with the corresponding Gaussian fits.
From the positions of the fits we extract a spin splitting for the first Landau level of $16.7~\text{meV}=|g^*|\mu_{\text{B}}B$ giving an effective Land\'e g-factor of $|g^*|=41$. From the mean difference of the first two Landau levels we get an effective mass of $m^*=(\hbar e B)/(37~\text{meV}) = 0.022 \times m_{0}$ ($m_{0}$: free electron mass). These values are in good agreement with previous STS measurements on InSb \cite{Hashimoto2008,Becker2010} and differ slightly from the known low temperature values for InSb at the conduction band minimum $m^{*}_{0}/m_0 = 0.0135$ \cite{Vurgaftman2001} and $g=-51$ \cite{Landolt} due to the well-known nonparabolicity of the conduction band of InSb.

The full width at half maximum (FWHM) of the fits decreases with Landau level index $n$, being $16.0~\text{mV}$ for $n=0$, $12.7~\text{mV}$ for $n=1$ and $12.1~\text{mV}$ for the first spin level of $n=2$. This effect is due to the decreased sensitivity of the Landau level drift states to the potential landscape, which can only probe potential fluctuations down to length scales of $r_{\text{c}}=\sqrt{(2n+1)\hbar/(|e|B)}$ \cite{Hashimoto2008}. Single spot $dI/dV$ spectra only show a width of about $11~\text{mV}$ for $n=0$. From this, we can estimate the width of the potential disorder to be $\Delta_{\text{dis}} \simeq \sqrt{16^2-11^2}~\text{meV}= 12~\text{meV}$.

The main cause for the potential disorder are the charged acceptors within the subband layer. The potential disorder of a highly doped \textit{p}-InSb(110) sample covered with Cs has been larger with $\Delta_{\text{dis}}\simeq 20$~meV \cite{Becker2010}. A previously analyzed 2DES prepared by Cs coverage on \textit{n}-InSb(110) showed a potential disorder $\Delta_{\text{dis}}$ comparable to our system with $\Delta_{\text{dis}}\simeq10\text{--}15$~meV, deduced from the lowest spin level at $T=300~\text{mK}$ and $B=12~\text{T}$ \cite{Hashimoto2008}. The disorder of a 2DES created by Fe adsorption on \textit{n}-InAs(110) showed a potential disorder of about $\Delta_{\text{dis}}\simeq 20~\text{meV}$ \cite{Morgenstern2002,Morgenstern2003-3}, which was determined by two independent methods: an STM image of the potential landscape was generated by probing the surface with a tip-induced quantum dot state \cite{Morgenstern2002} and FWHM of Landau level peaks measured at $T=6~\text{K}$ and $B=6~\text{T}$ were evaluated~\cite{Morgenstern2003-3}. However, these measurements are not spin resolved due to the lower $g$-factor of InAs, so a direct comparison of $\Delta_{\text{dis}}$ is difficult.

\begin{figure}\capstart
\includegraphics[width=8.5cm]{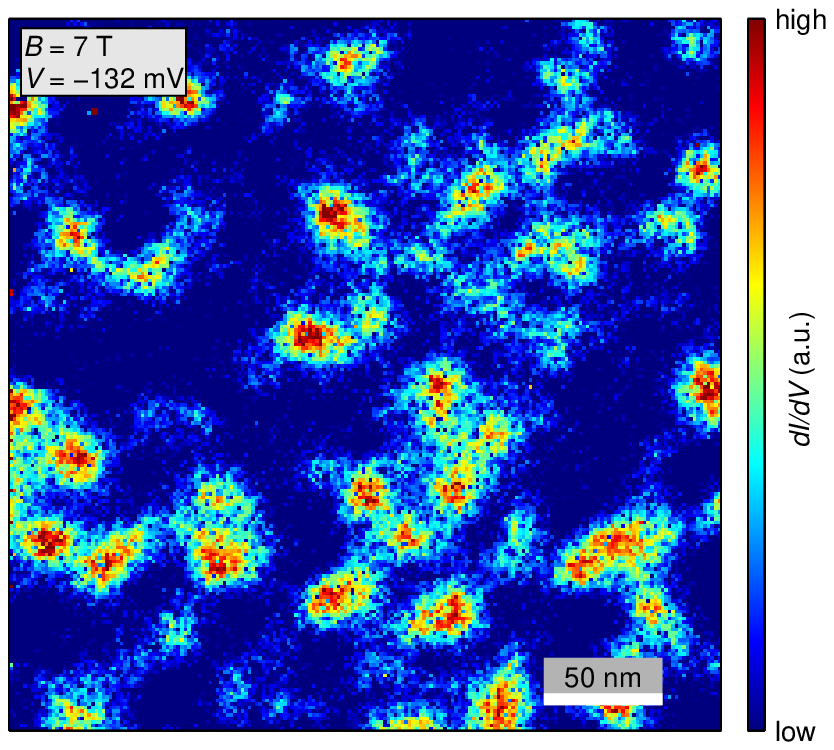}
\caption{$dI/dV(x,y)$ image at $V=-132~\text{mV}$ and $B = 7~\text{T}$, $V_{\text{stab}} = 300~\text{mV}$, $I_{\text{stab}} = 200~\text{pA}$, $V_{\text{mod}} = 1.6~\text{mV}_{\text{rms}}$, $300~\text{nm} \times 300~\text{nm}$. High $dI/dV$ intensity marks potential pits.\label{fig:Potential}}
\end{figure}

The lateral distribution of the potential disorder can be imaged by scanning tunneling spectroscopy (STS) in real space as is shown in Fig.~\ref{fig:Potential}.
The voltage of this $dI/dV$ image corresponds to the onset of the first Landau level and, thus, a high $dI/dV$ intensity exhibits the potential pits. There are roughly 40 local potential minima within the imaged area, giving an average distance of $\sqrt{(300~\text{nm})^2/40}\simeq 50~\text{nm}$ in accordance with visual inspection.

\section{Band bending and subband calculation}
It is well-known that submonolayer amounts of atoms adsorbed on an InSb surface lead to a large band bending near the surface. The adsorbed atoms act as donors to the underlying substrate with a donor level in the conduction band. For cesium on \textit{n}-InSb(110), this donor level has been determined by photoelectron spectroscopy to be $290~\text{meV}$ above the conduction band minimum \cite{Betti2001}. In our case of \textit{p}-doped InSb, where the bulk Fermi level $E_{\text{F}}$ lies at the valence band edge, we have to add the low temperature band gap of $E_{\text{gap}} = 235~\text{meV}$ \cite{Vurgaftman2001} to get an estimate of the total band shift at the surface of $V_{\text{bb}}=525~\text{meV}$. The corresponding band bending leads to electrons within an inversion layer which form two-dimensional subbands. The curvature of the potential inside the semiconductor, therefore, depends on the charge distribution within the inversion layer and on the density of charged acceptors.

The sample we used has an acceptor concentration of $N_{\text{A}} = 1.1 \times 10^{21}~\text{m}^{-3}$, which is orders of magnitude lower than the inversion layer electron concentration. To get an estimate of the expected subband energies without having to solve the Poisson equation and the full Hamiltonian including the nonparabolicity of InSb self-consistently, we approximate the inversion layer distribution by the 3D bulk density of states:
\begin{eqnarray}
  D_{\text{3D}}(E)=\frac{[2m^{*}(E)]^{3/2}}{2\pi^{2}\hbar^{3}}\sqrt{E}.
\end{eqnarray}
The energy dependence of the effective mass $m^{*}$ due to the nonparabolicity is approximated by $m^{*}(E)=m^{*}_{0} (1 + 2 E/E_{\text{gap}})$ \cite{Merkt1987} with the known effective mass at the conduction band minimum $m^{*}_{0}/m_0 = 0.0135$ \cite{Vurgaftman2001}. This results in a band bending as shown in Fig.~\ref{fig:dIdV_spatial}(a) of the main article with an inversion layer electron concentration of $N_{\text{s}} = 2.7 \times 10^{16}~\text{m}^{-2}$.

To obtain the subband positions $E_{i}$, $i\in\mathbb{N}_{0}$ within this potential, we use the triangular well approximation \cite{Ando1982} with electric fields approximated by the mean slope of the potential up to $E_{i}$ and an effective mass using the approximation $m^{*}(E)=m^{*}_{0} (1 + 2 (E_{i}/3+E_{\parallel})/E_{\text{gap}})$ \cite{Merkt1987}. Here, the energy $E$ splits up into the in-plane energy $E_{\parallel}$ and the energy $E_{i}$ at the onset of the subbands. The subband positions are marked in Fig.~\ref{fig:dIdV_spatial}(a) by horizontal lines. The density of subbands above the Fermi level is relatively large because of the flat band bending at a depth $z$ beyond the inversion layer, where only the acceptor density can screen the electric field. The first two subbands $E_{0}$ and $E_{1}$ are in good agreement with the experiment as shown in Fig.~\ref{fig:dIdV_spatial}(b).

\begin{figure}\capstart
\includegraphics[width=8.5cm]{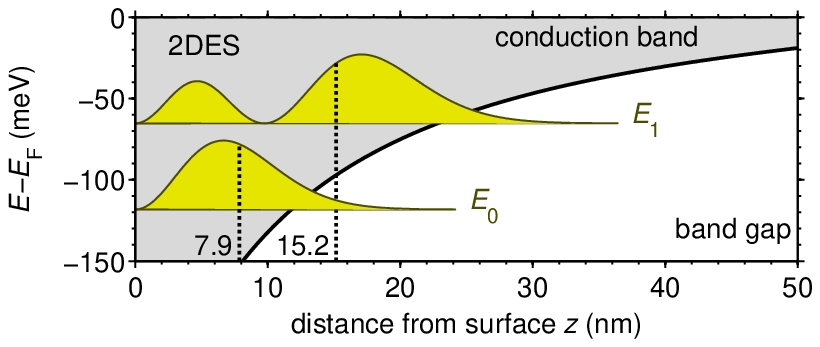}
\caption{Zoom of Fig.~\ref{fig:dIdV_spatial}(a): Calculated band bending (black line) and the first two 2D subbands with their electron distribution curves $|\psi_{i}(z)|^{2}$ (yellow areas). The mean depths $\left\langle{z}\right\rangle$ of each distribution is indicated by the dotted lines and marked in nm.\label{fig:PoissonS}}
\end{figure}

The 2D electron concentrations of the first two subbands are $N_{0} = 1.3 \times 10^{16}~\text{m}^{-2}$ and $N_{1} = 0.7 \times 10^{16}~\text{m}^{-2}$. Their electron distributions along $z$ are plotted in Fig.~\ref{fig:PoissonS}.
From these distributions, the mean depth of the subband electrons below the surface is deduced to be $7.9~\text{nm}$ for the first subband and $15.2~\text{nm}$ for the second subband.

The averaged $dI/dV$ curve (curve (II) of Fig.~\ref{fig:dIdV_spatial}(b)) exhibits rounded steps close to the energies calculated by the Poisson-Schr{\"o}dinger equation. The width of these steps is caused by disorder. Indeed, spatially resolved $dI/dV$ images as shown e.g.\ in Fig.~3 of Ref.~\cite{Morgenstern2002} reveal quantum dot like states at low energy which percolate close to the onset of the plateau. The decreasing height of the step with increasing subband index is consistently observed for different 2DES \cite{Morgenstern2002, Wiebe2003, Becker2010} and is attributed to the fact that higher subbands are reaching deeper into the crystal and, thus, have a lower LDOS at the surface. This is clearly visible in Fig. \ref{fig:PoissonS}.

\section{Coulomb gap and energy resolution}
In order to account for the finite temperature $T=5~\text{K}$ and for screening effects by the metallic STM tip, we have modified the V-shaped DOS of Eq.~(\ref{eq:CoulombGap}) of the main article called $D_{0}(E)$ by a finite DOS at $E_{\text{F}}$ according to Ref.~\onlinecite{Pikus1995}:
\begin{equation}
D_{\text{mod}}(E_{\text{F}})\approx 0.085 \frac{4\pi\varepsilon_{0}\varepsilon_{r}}{e^2d}+0.86\frac{(4\pi\varepsilon_{0}\varepsilon_{\text{r}})^2k_{\text{B}}T}{e^4}
\end{equation}
with $d$ being the distance of a metallic gate to the center of the 2DES and $k_{\text{B}}$ being the Boltzmann constant. In our system, $d$ can be estimated by the mean depth of the first subband as calculated above and the distance of the STM tip to the surface. The distance of the tip to the surface is estimated by fitting $I(z)$ measurements to an exponential tunneling decay and extrapolating the $z=0$ offset to the quantum of conductance $G_0=2e^2/h$ with electron charge $e$ and Planck's constant $h$ \cite{Kroeger2007}. From this, we get a tip distance of about $700~\text{pm}$ at the stabilizing current of this measurement $I_{\text{stab}}=400~\text{pA}$. This results in $d=8.6~\text{nm}$. With $\varepsilon_{\text{r}}=16.8$ \cite{Dixon1980}, we get $D_{\text{mod}}(E_{\text{F}})=1.66\times10^{17}~\text{eV}^{-1}\text{m}^{-2}$. For the sake of simplicity, we simply cut off the V-shaped Coulomb gap
$D_{0}(E)$ by this value leading to $D(E) =\max{\{D_{0}(E),D_{\text{mod}}(E_{\text{F}})\}}$.

The Coulomb gap $D(E)$ is finally broadened by the energy resolution of the STS experiment being $\delta E \simeq \sqrt{(3 k_{\text{B}} T)^2 + (2.5 V_{\text{mod}})^2}=4.2~\text{meV}$ \cite{Wachowiak,Haude} with $V_{\text{mod}}=1.6~\text{mV}_{\text{rms}}$ and $T=5~\text{K}$. In order to calculate the resulting $dI/dV$ curve, we approximate a Gaussian energy broadening with $\delta E$ being twice its standard deviation. This Gaussian is folded with the calculated $D(E)$ resulting in the red dotted curve of Fig. 4(a) of the main article. Notice that no fit parameter is involved within this calculation which excellently reproduces the measured Coulomb gap.

\section{Local Coulomb gap within the potential landscape}
The Coulomb gap in the DOS has been measured and discussed within the main article via spatially averaged $dI/dV$ spectra. The main article and the attached video file shows, in addition, that single $dI/dV$ peaks measured at a fixed position also exhibit a Coulomb gap, i.e.\ a reduction of intensity while moved across $E_{\text{F}}$ by $B$ field. Such a Coulomb gap can also be identified at fixed magnetic field in local spectra taken at different positions within the potential landscape.

\begin{figure}\capstart
\includegraphics[width=8.5cm]{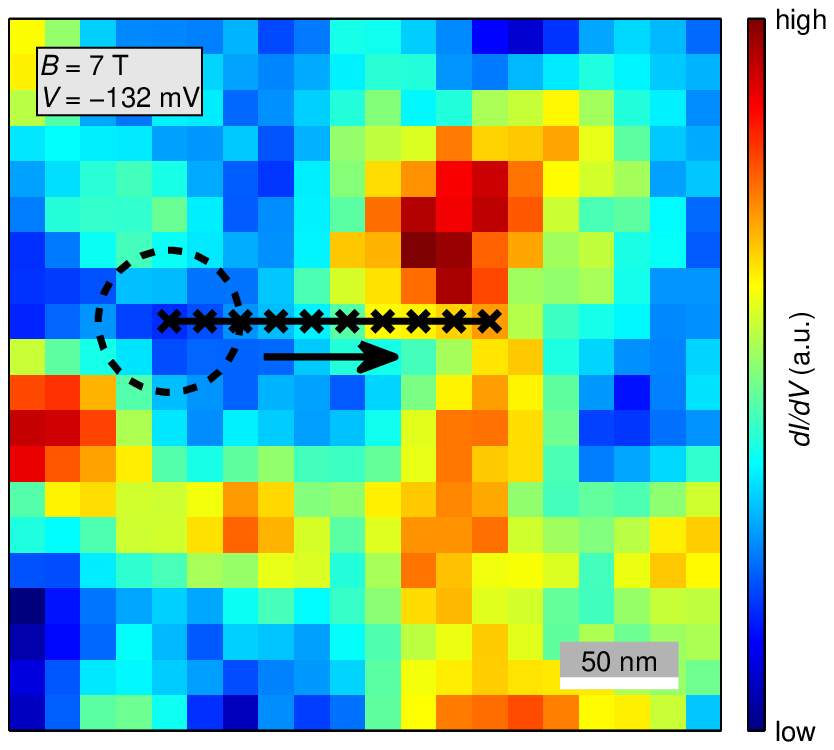}
\caption{$dI/dV(x,y)$ image at $V=-132~\text{mV}$ and $B = 7~\text{T}$; the data belong to the spatially averaged $dI/dV$ curve in Fig.~\ref{fig:dIdV_spatial}(b)(III) of the main article; $V_{\text{stab}} = 300~\text{mV}$, $I_{\text{stab}} = 200~\text{pA}$, $V_{\text{mod}} = 0.4~\text{mV}_{\text{rms}}$, $300~\text{nm} \times 300~\text{nm}$, $20\times20$ spectra. High $dI/dV$ intensity marks potential pits as in Fig.~\ref{fig:Potential}. The crosses and the arrow mark the positions and the sequence of the spectra shown in Fig.~\ref{fig:LCEinzel}. 12 next neighboring spectra have been averaged for each spectrum in order to reduce noise. This leads to a spatial resolution of 60~nm as indicated by the circle. \label{fig:LokalCoulomb}}
\end{figure}
\begin{figure}\capstart
\includegraphics[width=8.5cm]{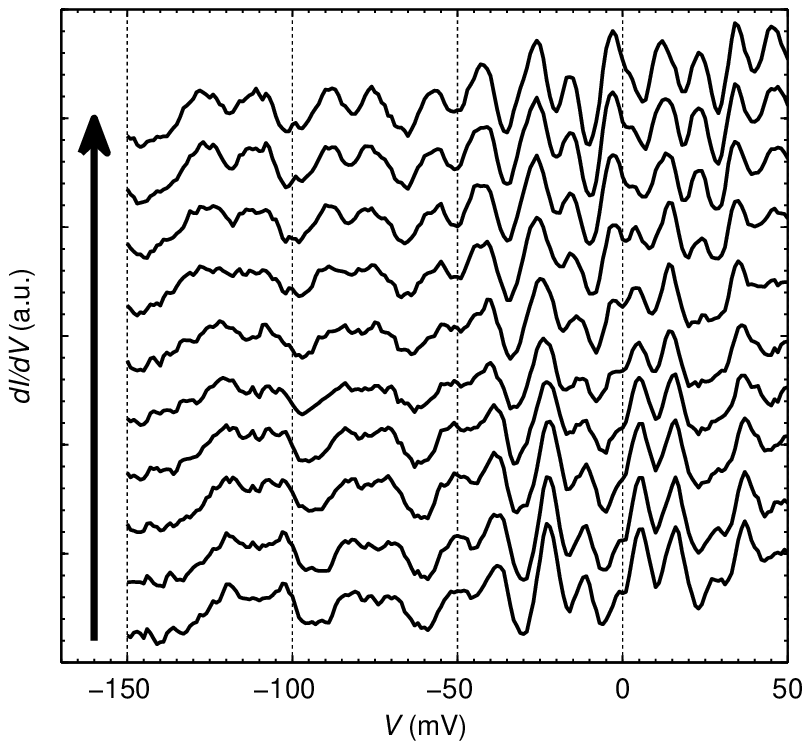}
\caption{$dI/dV$ spectra recorded at different positions as marked in Fig.~\ref{fig:LokalCoulomb}. The spectra are offset vertically. The Coulomb gap is visible as a dip or kink pinned at $V=0~\text{mV}$ ($E_{\text{F}}$). Note that all other features except the Coulomb gap feature shift continuously due to the varying potential.\label{fig:LCEinzel}}
\end{figure}

Fig.~\ref{fig:LokalCoulomb} shows a low energy $dI/dV$ image marking the potential disorder of the same area as imaged in Fig.~\ref{fig:Potential}. Fig.~\ref{fig:LokalCoulomb} has lower spatial resolution, but higher energy resolution. The averged spectrum of all spectra gives the DOS of Fig.~\ref{fig:dIdV_spatial}(b)(III) of the main article. Each pixel represents a spatial average of 12 neighboring spectra, thus exhibiting the averaged potential of an area of about $50~\text{nm}^2$. The spectra corresponding to the marked positions are shown in Fig.~\ref{fig:LCEinzel}.

Moving down in the potential landscape as indicated by the arrow, the peaks belonging to the three lowest Landau levels shift to lower energies as expected. Near the Fermi level, additional subbands lead to a more irregular pattern, but the Landau levels of the lowest subband still dominate the pattern. If observed carefully, all features shift to lower energies, except the suppression caused by the Coulomb gap, which is pinned to $V=0~\text{mV}$. This suppression is visible as a dip at medium potential, where a spin state is exactly at $E_{\text{F}}$, and as kinks in the tails of the peaks at low and high potential. This plot illustrates again the local observation of a Coulomb gap feature, now by averaging over distances belonging to a single potential minimum or maximum. Thus, the Coulomb gap appears to be ubiquitous even in local spectra, which is not expected from the typical qualitative arguments.

\section{Theoretical estimate of the exchange enhancement}
In this Section, we briefly sketch how to obtain a quantitative estimate of the filling factor dependence of the effective $g$-factor resulting from the Coulomb interaction. We start out by introducing the Hamiltonian describing a two-dimensional free electron gas subject to a perpendicular magnetic field $B$ (pointing in $z$-direction):
\begin{equation}\label{h0}
H_0 = \sum_{nk\sigma}\epsilon_{n\sigma} c_{nk\sigma}^\dagger c_{nk\sigma}^{\phantom{\dagger}}~,
\end{equation}
with $c_{nk\sigma}$ being an annihilation operator in second quantization. The single-particle energies read
\begin{equation}
\epsilon_{n\sigma} = \hbar\omega_c\left(n+\frac{1}{2}\right) + g^* \frac{\sigma}{2}\frac{e\hbar}{2m_0} \frac{B}{c}~,~~~\omega_c=\frac{eB}{m^*c}~,
\end{equation}
where the effective mass $m^*=0.02m_0$ and effective $g$-factor $g^*=-42$ can both be extracted from our experimental data. We have chosen the vector potential $\vec A(\vec x) = Bx\vec e_y$ in the Landau gauge, and the $n$-th Landau level of electrons with spin $\sigma=\pm$ is thus characterized by an additional quantum number $k$ associated with the momentum operator $\hat p_y$ of the $y$-direction. In contrast, the interaction is naturally expressed in terms of \textit{two-dimensional} momentum quantum numbers $\vec k=(k_x,k_y)$:
\begin{equation}\label{ww}
U = \frac{1}{2\mathcal V}\sum_{\stackrel{\vec k_1 \vec k_2}{\vec q\neq0}}\sum_{\sigma_1\sigma_2} V(|\vec q|)
 c_{\vec k_1+\vec q \sigma_1}^\dagger c_{\vec k_2-\vec q \sigma_2}^\dagger
c_{\vec k_2\sigma_2}^{\phantom{\dagger}} c_{\vec k_1\sigma_1}^{\phantom{\dagger}}~.
\end{equation}
Using cgs units, the Fourier transform of the Coulomb potential yields
\begin{equation}
V(q) = F(q)\frac{2\pi e^2}{\varepsilon_{\text{r}}q}~,
\end{equation}
with the form factor $F(q)$ accounting for the finite thickness of the inversion layer \cite{Ando1974}:
\begin{equation}
F(q) = \frac{3}{8x}+\frac{3}{8x^2}+\frac{1}{4x^3}\,,~x=1+\frac{q}{\sqrt[3]{\frac{48\pi m^*e^2}{\varepsilon_{\text{r}}\hbar^2}\left(\frac{11}{32}N_{\text{s}}+N_{\text{d}}\right)}}~.
\end{equation}
The dielectric constant $\varepsilon_{\text{r}}$ of InSb is given by $\varepsilon_{\text{r}}=16.8$, and $N_{\text{s}}=2.7 \times 10^{16}~\text{m}^{-2}$ as well as $N_{\text{d}}=8 \times 10^{14}~\text{m}^{-2}$ are the total electron density in the inversion layer as well as the density of the ionized acceptors, respectively.

In order to compute the self-energy $\Sigma$ of the many-particle problem posed by the Hamiltonian $H=H_0+U$, we employ the random phase approximation (RPA) \cite{Ando1974}, which is reasonable for the problem at hand (at least for integer filling factors where the fractional quantum Hall effect is absent) since the density of the electrons occupying the lowest subband $N_{0}\approx25/a_0^2$ is large compared to the scale set by the Bohr radius $a_0=\hbar^2\varepsilon_{\text{r}}/m^*e^2$. The first (Fock exchange \cite{Janak1969}) term of the RPA series reads
\begin{widetext}\begin{equation}\label{se}\begin{split}
\Sigma_{n_1k_1,n_2k_2}^\sigma(i\omega) & = -T \sum_{n_3k_3}\sum_{i\Omega}\,
\langle n_1k_1\sigma,n_3k_3\sigma|U|n_3k_3\sigma,n_2k_2\sigma\rangle\, \langle n_3k_3\sigma|\mathcal G_0(i\omega+i\Omega)|n_3k_3\sigma\rangle \\
&=-T\delta_{n_1n_2}\delta(k_1-k_2)\int\frac{qdq}{2\pi}\sum_{n_3} P_{n_1n_3}\left(q^2l^2/2\right)
\sum_{i\Omega}\frac{V(q)}{i\omega+i\Omega-\epsilon_{n_3\sigma}+\mu}~,
\end{split}\end{equation}\end{widetext}
where $\mathcal G_0(i\omega)$ is the noninteracting Matsubara Green function related to $H_0$, $l=\sqrt{\hbar c/eB}$ the magnetic length, and $T=5~\text{K}$ as well as $\mu=\nu\hbar\omega_c/2$ denote the temperature and the chemical potential, respectively. We have defined the following combination of associated Laguerre polynomials $L_{n_1}^{n_2-n_1}(x)$:
\begin{equation}
P_{n_1n_2}(x) = (-1)^{n_1+n_2}e^{-x}L_{n_2}^{n_1-n_2}(x) L_{n_1}^{n_2-n_1}(x)~.
\end{equation}
In order to derive the second equality of Eq.~(\ref{se}), one needs to evaluate the two-particle matrix element $\langle n_1k_1\sigma,n_3k_3\sigma|U|n_3k_3\sigma,n_2k_2\sigma\rangle$ of the Coulomb interaction. This can be achieved by resorting to the momentum representation of the Landau states $|nk\sigma\rangle$,
\begin{equation}\begin{split}
\langle k_xk_y\sigma|nk\sigma\rangle & = \delta(k-k_y) (-i)^n e^{ik_xk_yl^2}\\
& ~~~~\times\sqrt{\frac{l}{\sqrt{\pi}n!2^n}}\,e^{-k_x^2l^2/2} H_n(lk_x)~,
\end{split}\end{equation}
inserting unit operators, and eventually carrying out an integral over products of Hermite polynomials $H_n(x)$ \cite{Gradstein1981}.

\begin{figure}[b]\capstart
\includegraphics[width=0.9\linewidth]{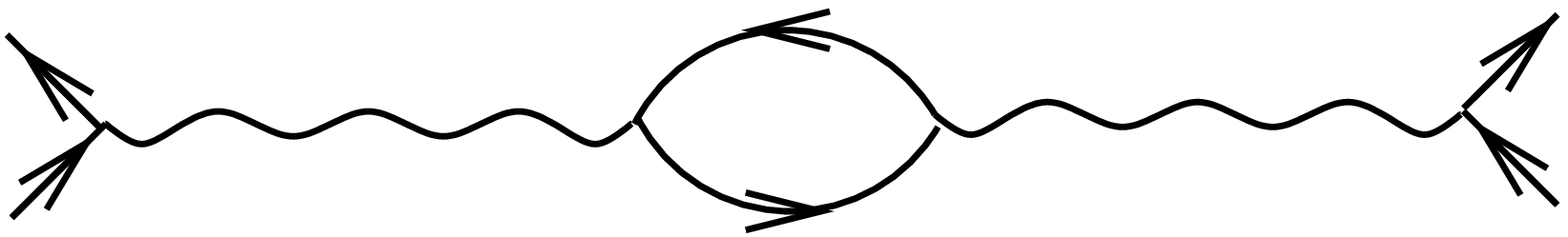}
\caption{Structure of the bubble diagram corresponding to the analytic expression of Eq.~(\ref{bubble}).}
\label{bubblefig}
\end{figure}

The second term of the RPA series involves the bubble diagram depicted in Fig.~\ref{bubblefig}. The corresponding analytic expression reads
\begin{widetext}\begin{equation}\label{bubble}\begin{split}
& -T V(\vec q) V(\vec q') \sum_{\vec k\vec k'\sigma}\sum_{i\lambda}
\big\langle k_x'k_y'\sigma \big|\mathcal G_0(i\lambda)\big|k_xk_y\sigma\big\rangle\,
\big\langle (k_x+q_x)(k_y+q_y)\sigma \big|\mathcal G_0(i\lambda+i\Omega)\big|(k_x'+q_x')(k_y'+q_y')\sigma\big\rangle \\
= & - V(\vec q)^2\delta(\vec q-\vec q')\frac{1}{2\pi l^2}\sum_{n_1n_2}P_{n_1n_2}\big(|\vec q|^2l^2/2\big)
\sum_\sigma \frac{f(\epsilon_{n_1\sigma}) - f(\epsilon_{n_2\sigma})}{i\Omega+\epsilon_{n_1\sigma}-\epsilon_{n_2\sigma}}
= - V(\vec q)^2\delta(\vec q-\vec q') \Pi_0(|\vec q|,i\Omega)~,
\end{split}\end{equation}\end{widetext}
with $f(\epsilon) = 1/\{1+\exp[(\epsilon-\mu)/T]\}$ being the Fermi function. The second equality can again be established by inserting unit operators in order to calculate the (two-dimensional) momentum matrix elements of the noninteracting Green function. Since Eq.~(\ref{bubble}) is of the same (momentum-conserving) structure as the Coulomb potential, the geometric RPA series can be summed up in complete analogy with the well-known case where the magnetic field is absent \cite{Bruus2004}. The resulting self-energy is eventually given by Eq.~(\ref{se}) with the bare $V(q)$ replaced by
\begin{equation}
V_{\text{RPA}}(q,i\Omega) = \frac{V(q)}{1-V(q)\Pi_0(q,i\Omega)}~,
\end{equation}
which we moreover set to its zero-frequency value $i\Omega=0$ \cite{Ando1974}. The remaining Matsubara sum in Eq.~(\ref{se}) can then be carried out analytically, and the self-energy (which is frequency-independent, ruling out difficulties with the analytic continuation to the real axis from the beginning) can be computed with minor numerical effort. Surprisingly, it turns out that such a static screening approximation is sufficient to reproduce the results of more elaborate approaches \cite{Smith1992} on a quantitative level, providing the \textit{a posteriori} motivation to solely stick to this scheme.

Since the self-energy is diagonal \cite{MacDonald1986} w.r.t.~$n$ and $k$, the magnitude of the oscillations shown in Fig.~\ref{fig:CoulombGap}(c) of the main text is given by
\begin{equation}
\delta(\nu) = \left(\Sigma^\downarrow_{n=0}-\Sigma^\uparrow_{n=0}\right)_\nu - \left(\Sigma^\downarrow_{n=0}-\Sigma^\uparrow_{n=0}\right)_{\nu-1}~,
\end{equation}
where the two bracketed terms are to be computed at filling factors $\nu$ and $\nu-1$, respectively. The result is $\delta(11)=0.76~\text{meV}$, $\delta(15)=0.55~\text{meV}$ in nice agreement with the experimental data. We emphasize that these values only change slightly when the physical system parameters (such as the dielectric constant, the temperature, the effective $g$-factor, or the effective mass) are varied within reasonable limits. E.g., using $m^*=0.04m_0$ instead of $m^*=0.02m_0$ leads to $\delta(11)=0.82~\text{meV}$, $\delta(15)=0.59~\text{meV}$.

\end{document}